\documentclass[twocolumn]{aastex63}
\shorttitle{Supermassive Black Hole Binary in NGC\,4472}
\shortauthors{Wrobel \& Lazio}

\begin{document}

\title{Toward Astrometric Constraints on a Supermassive Black Hole Binary\\
in the Early-Type Galaxy NGC\,4472}

\author[0000-0001-9720-7398]{J. M. Wrobel}
\affiliation{National Radio Astronomy Observatory, P.O. Box O,
  Socorro, NM 87801, USA}

\author{T. J. W. Lazio}
\affiliation{Jet Propulsion Laboratory, California Institute of
  Technology, 4800 Oak Grove Drive, Pasadena, CA 91109, USA}

\correspondingauthor{J. M. Wrobel}
\email{jwrobel@nrao.edu}

\received{2022 February 8}\accepted{2022 April 17}
\submitjournal{ApJ}

\begin{abstract}
The merger of two galaxies, each hosting a supermassive 
black hole (SMBH) of mass $10^6$\,M$_{\odot}$ or more, could yield a
bound SMBH binary. For the early-type galaxy NGC\,4472, we study how
astrometry with a next-generation Very Large Array (ngVLA) could be 
used to monitor the reflex motion of the primary SMBH of mass 
$M_{\rm pri}$, as it is tugged on by the secondary SMBH of mass 
$M_{\rm sec}$. Casting the orbit of the putative SMBH binary in terms
of its period $P$, semimajor axis $a_{\rm bin}$, and mass ratio 
$q = M_{\rm sec} / M_{\rm pri} \le 1$, we find the following: (1)
Orbits with fiducial periods of $P = 4$\,yr and 40\,yr could be
spatially resolved and monitored. (2) For a 95\% accuracy of 
$2\,\mu$as per monitoring epoch, sub-parsec values of $a_{\rm bin}$
could be accessed over a range of mass ratios notionally encompassing
major ($q > \frac{1}{4}$) and minor ($q < \frac{1}{4}$) galaxy mergers.
(3) If no reflex motion is detected for $M_{\rm pri}$ after 1(10)\,yr 
of monitoring, a SMBH binary with period $P = 4(40)$\,yr and mass ratio
$q > 0.01(0.003)$ could be excluded. This would suggest no present-day
evidence for a past major merger like that recently simulated, where 
scouring by a $q \sim 1$ SMBH binary formed a stellar core with 
kinematic traits like those of NGC\,4472. (4) Astrometric monitoring 
could independently check the upper limits on $q$ from searches for 
continuous gravitational waves from NGC\,4472.
\end{abstract}


\section{Motivation}

The merger of two galaxies, each hosting a supermassive black hole 
(SMBH) of mass $10^6$\,M$_{\odot}$ or more, is expected to yield an 
SMBH binary. Eventually the orbit of the SMBH binary will shrink 
due to gravitational wave (GW) emission \citep{beg80}. Such GW 
signatures are currently being sought with the Parkes Pulsar Timing
Array \citep[PPTA;][]{man13,sha15,las16}, the European Pulsar 
Timing Array \citep{len15,bab16,des16}, and the North American 
Nanohertz Observatory for Gravitational Waves 
\citep[NANOGrav;][]{arz18}. It is also anticipated that such GW 
signatures will be detected with the future Laser Interferometer 
Space Array \citep{ama17}.

In addition, searches are underway for the electromagnetic (EM) 
signatures of SMBH binaries. Decades of data seeking indirect 
EM signatures, such as periodicities in the optical emission-line 
velocities or photometric variability of active galactic nuclei 
(AGNs), have been used to identify candidate SMBH binaries. See 
\citet{bur19} and \citet{der19} for recent reviews of all of the 
topics mentioned above.

One direct EM signature would be to spatially resolve and monitor
the orbit of one or both members of an SMBH binary. \citet{ban17} 
report possible evidence for orbital motion in the 7.3-pc binary in 
the radio galaxy 0402+379. But the estimated period spans millennia,
making it difficult to acquire sufficient observables to solve for an 
orbit. EM strategies for studying tighter and more easily measured
orbits are only now being devised \citep{dor18,saf19,dex20}. These
strategies rely upon recent or projected advances in interferometric
techniques at millimeter (mm) or near-infrared wavelengths.

Here, we consider a strategy proposed by \citet{saf19} for continuum 
targets at Jansky (Jy) levels observable at 230\,GHz (1.3\,mm) with the
Event Horizon Telescope \citep[EHT;][]{eht19}, and adapt it for the 
fainter, more abundant targets at milliJansky (mJy) levels that could
be observed at 80\,GHz (3.7\,mm) with a next-generation Very Large
Array\footnote{https://ngvla.nrao.edu} \citep[ngVLA;][]{mur18}.

Section~2 describes the selection of an example target, while
Section~3 describes an observing strategy that capitalizes on a
powerful, designed-in capability of the ngVLA at its highest
resolutions, namely paired antenna calibration
\citep{car99,car21}. Section~4 explores the implications of
high-resolution ngVLA observations for astrometric monitoring and for
related tie-ins to galaxy evolution and to multimessenger astronomy in
the context of GW observations. We close in Section~5 with a summary
and conclusions. A preliminary version of this work appeared in
\citet{wro21}.

\section{Target Selection}

We focus on the early-type galaxy NGC\,4472, the dominant galaxy in
Virgo Subcluster~B \citep[e.g.,][]{jan10,arr12}. Throughout we assume
a distance of 16.7\,Mpc, where 81.0\,pc subtends 1\arcsec\, \citep{bla09}.

NGC\,4472 may have experienced one or more galaxy mergers, as
suggested from its overall slow rotation and its stellar core, which
exhibits a substantial mass deficit, counter rotation and
tangentially-biased orbits
\citep[e.g.,][]{bet01,kra11,ems11,tho14}. Such core traits are linked
to stellar scouring by an SMBH binary, a potential byproduct of a
galaxy merger \citep[e.g.,][]{mil02,gua08,kor09}.  Finally, though
somewhat fortuitous, \citet{min17} conduct a series of simulations to
construct realizations of the local GW landscape. They find that
NGC\,4472 plausibly could host a SMBH binary and illustrate some of
their results by considering it specifically.

High-resolution imaging of the low-luminosity AGN in NGC\,4472 using 
the technique of very long baseline interferometry (VLBI) is available
only at 5\,GHz (6.0\,cm) and 8.4\,GHz (3.6\,cm) \citep{and05,nag05}. 
The most constraining size information comes from the Very Long 
Baseline Array image at 8.4\,GHz showing a 4-mJy source \citep{and05}.
The diameter of the cm source has a full width at half maximum (FWHM)
of less than 0.73\,mas (0.059\,pc). Attempts to probe smaller size
scales, via time variability at 8.4\,GHz and 15\,GHz (2.0\,cm) using
the Very Large Array, were inconclusive \citep{and05,nag05}. 

Hydrodynamical simulations including radiative processes and SMBH 
feedback suggest that NGC\,4472's low-luminosity AGN can be 
understood as arising from jet-like outflows traced via synchroton 
emission and launched from a radiatively-inefficient accretion 
inflow \citep[e.g.,][]{ina20}.


\begin{figure}[!t]
\centering
\includegraphics[angle=0,scale=0.4]{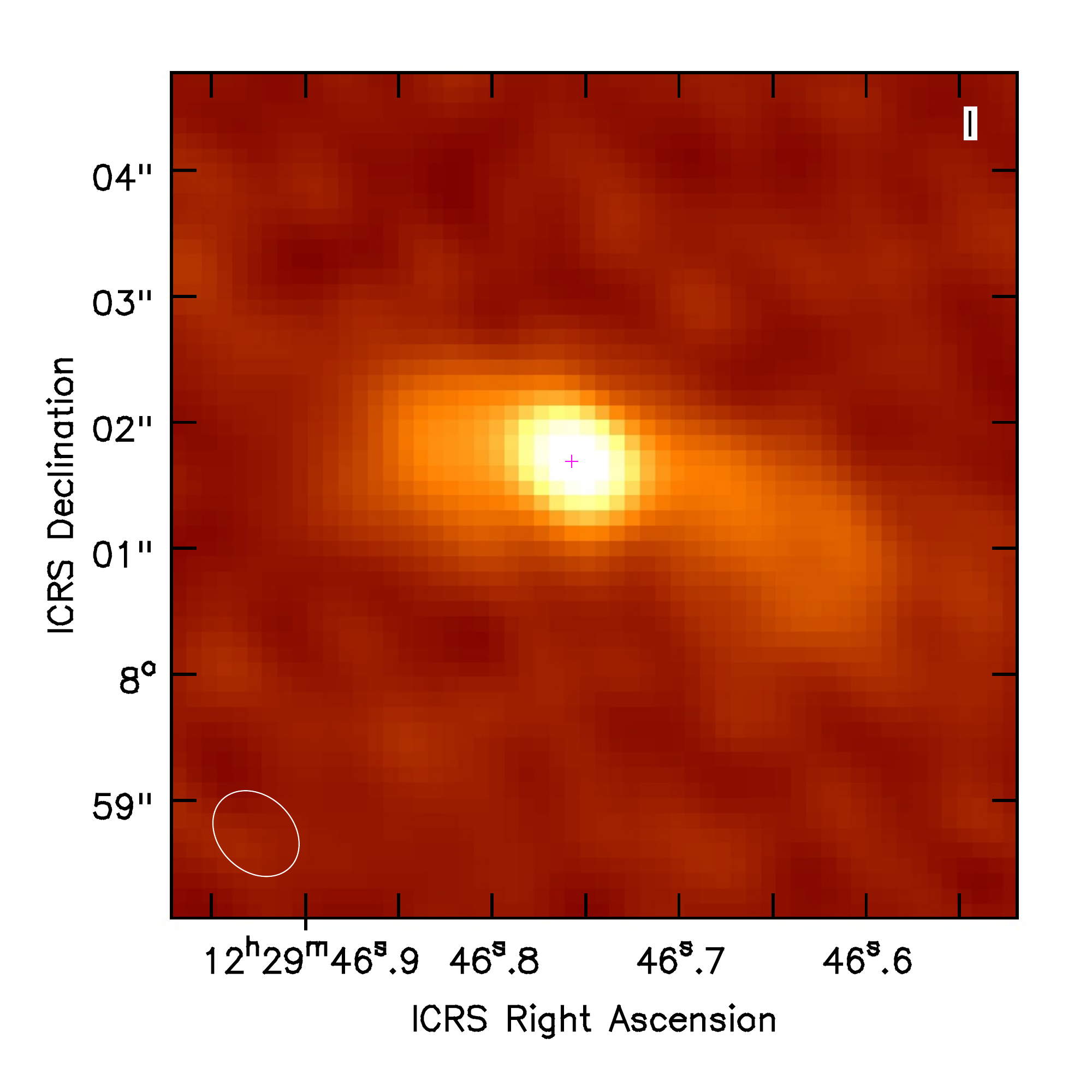}
\caption{ALMA archival image of the emission at 98\,GHz (3.1\,mm) from
  the low-luminosity AGN in NGC\,4472. The scale is 1\arcsec\, =
  81.0\,pc \citep{bla09}. The ellipse in the southeast corner shows
  the synthesized beam dimensions at FWHM of 0\farcs76 (62\,pc)
  $\times$ 0\farcs59 (48\,pc) with an elongation position angle of
  46\arcdeg. The cross marks the location of the emission peak, which
  has a value of 2.92\,mJy\,beam$^{-1}$. The white box in the
  northwest corner highlights the label $I$ for the Stokes parameter.}
\end{figure}

\section{ngVLA Observations}

We adopt the ngVLA's Long Baseline Array (LBA) component, which 
features ten sites and three 18~m antennas per site \citep{car21}. We
assume an observing frequency of 80\,GHz to match that assigned to 
notional continuum studies in the highest frequency band \citep{wro20}.
The point spread function after tapering the natural weights is
modelled\footnote{https://gitlab.nrao.edu/vrosero/ngvla-sensitivity-calculator}
to have a FWHM of $PSF = 0.1$\,mas (0.0081\,pc).

To accrue a reasonable time on the target, one antenna per LBA site 
would observe the phase calibrator while the other two continuously
observed NGC\,4472 \citep{car99,car21}. This capability is generally
not available with current VLBI arrays, but has been  designed into
the ngVLA's LBA component. After 4~hours on target with 
dual-polarization receivers and a bandwidth of 20\,GHz per 
polarization, the thermal noise is modelled$^2$ to have a 
root-mean-square value of about 3\,$\mu$Jy\,beam$^{-1}$.

The overall astrometric accuracy of the ngVLA observation will have
contributions from terms due to (i) the signal-to-noise ratio $S/N$ on
the target, $\sigma_{\rm s/n}$; and (ii) the relative accuracy
achieved via phase referencing, $\sigma_{\rm pr}$. Regarding term~(i),
Figure~1 shows an image of NGC\,4472 at 98\,GHz (3.1\,mm), obtained
from the archives of the Atacama Large Millimeter/submillimeter Array
(ALMA) using the Cube Analysis and Rendering Tool for Astronomy
(CARTA).\footnote{https://doi.org/10.5281/zenodo.4905459} If the peak
signal in Figure~1 is available to the ngVLA at 80\,GHz, it would have
a $S/N \sim 970$ and be localized with an associated accuracy of
$\sigma_{\rm s/n} = PSF / (1.665 \times S/N) \sim 0.1\,\mu$as. This
expression stems from equation (25) in \citet{con98} and applies to
Gaussian-based fitting.

\begin{figure*}[!t]
\centering
\includegraphics[angle=0,scale=0.8]{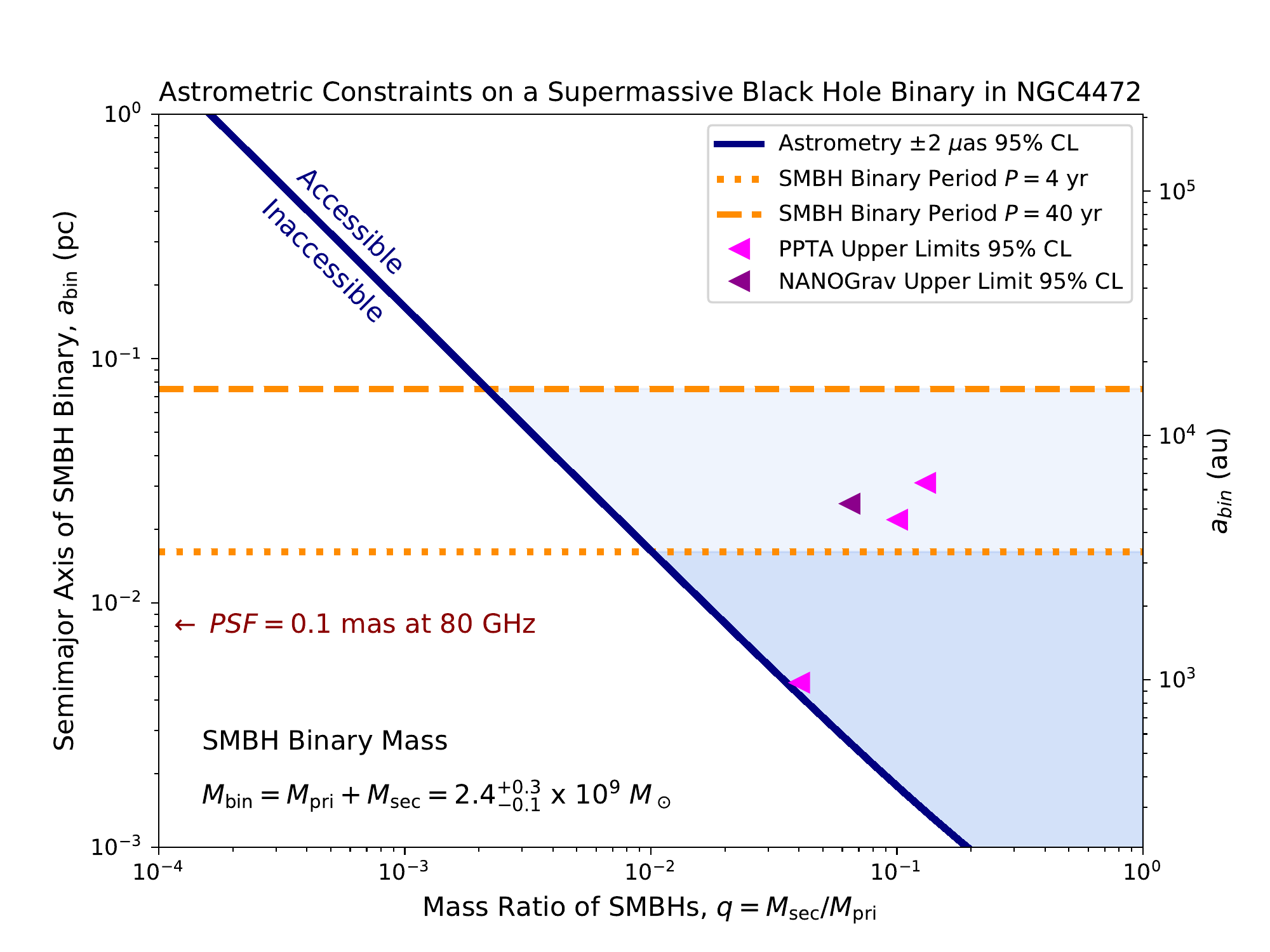}
\caption{Parameter space for $a_{\rm bin}$ and $q$ for a putative SMBH
  binary in the early-type galaxy NGC\,4472. $M_{\rm bin}$ is from
  \citet{rus13}. The region to the right of the navy diagonal line is
  accessible with ngVLA astrometric monitoring at 80\,GHz with the
  labelled accuracy. The associated $PSF$ of the ngVLA Long Baseline
  Array (LBA) is marked for reference. The GW constraints from the
  PPTA are tabulated in \citet{sch16}, while that from NANOGrav is
  derived from \citet{arz21}. All quantities invoke the \citet{bla09}
  distance of 16.7\,Mpc.}
\end{figure*}

Regarding term~(ii), the phase referencing will be imperfect due to
the different atmospheric conditions associated with two separations,
that between the sky locations of the target and the phase calibrator,
and that between the ground locations of the target antenna and the
calibrator antenna \citep[equation 13.128 of][]{tho17}. To reach
levels of $\sigma_{\rm pr} \sim 1\,\mu$as at mm wavelengths, it is
desirable to employ multiple phase calibrators whose sky locations are
separated from the target by less than 1-2\arcdeg\,
\citep{bro11,rei14,rio20}.  Several such calibrator candidates are
already known for NGC\,4472, one with a sub-degree
separation.\footnote{http://www.vlba.nrao.edu/astro/calib/} Some of
the calibrator candidates appear to be adequately strong according to
the ALMA Calibrator Source
Catalog.\footnote{https://almascience.nrao.edu/sc/} Others could be
briefly observed to assess their suitability. It is also desirable to
closely pack the target and calibrator antennas per LBA site. Below,
we assume that the overall astrometric accuracy of the ngVLA
observation will be dominated by term~(ii), with $\sigma_{\rm pr} \sim
1\,\mu$as.

\section{Implications}

\subsection{Astrometric Monitoring}

Following \citet{saf19}, we examine how astrometric monitoring of
NGC\,4472 could constrain the reflex motion of the low-luminosity
AGN's primary SMBH of mass $M_{\rm pri}$, as it is tugged on by a
putative secondary SMBH of mass $M_{\rm sec}$. These masses define a
binary mass $M_{\rm bin} = M_{\rm pri} + M_{\rm sec}$ and mass ratio
$q = M_{\rm sec} / M_{\rm pri} \le 1$. We allow $q$ to vary.  We adopt
$M_{\rm bin} = 2.4^{+0.3}_{-0.1} \times 10^9$\,M$_{\odot}$ from
\citet{sch16}, which is the dynamically derived mass estimate from
\citet{rus13} but linearly scaled to an assumed distance of 16.7\,Mpc
\citep{bla09}.  Because early-type galaxies with larger SMBH masses
are more likely to exhibit nuclear radio emission
\citep[e.g.,][]{nyl16}, we assume that the mm emission from NGC\,4472
is associated with its primary SMBH.

We assume a circular orbit for the SMBH binary and use Kepler's third
law to link the binary's semimajor axis $a_{\rm bin}$ to its mass 
$M_{\rm bin}$ and orbital period $P$ \citep[equation (1) of][]{dex20}.
The reflex motion of $M_{\rm pri}$ as it orbits, with semimajor axis
$a_{\rm pri}$, about the binary's center of mass is $a_{\rm pri} = 
a_{\rm bin} \times q / (1+q)$ \citep[equation (1) of][]{saf19}. We 
recast this as
\begin{equation}
a_{\rm bin} = a_{\rm pri} \times (1+q) / q. \label{(1)}
\end{equation}

Suppose that ngVLA astrometry of the low-luminosty AGN could achieve a
95\% accuracy of $2\,\mu$as for each epoch in a monitoring
sequence. Then the reflex constraint per epoch would become $a_{\rm
  pri} = 2\,\mu{\rm as} \times 81.0\,{\rm pc} / 10^6\,\mu{\rm as} =$
0.00016\,pc. Inserting this value into equation (1) then defines how
$a_{\rm bin}$ can be related to $q$. This behavior is shown as the
navy diagonal line in Figure~2. The parameter space to the right of
this line is accessible via ngVLA monitoring of NGC\,4472 with the
adopted astrometric accuracy.  The parameter space to the left of this
line is inaccessible.

A possible complication regarding the ngVLA monitoring deserves
mention: \citet{su19} suggest that NGC\,4472 is moving northward, 
relative to its surrounding X-ray-emitting gas, with a velocity 
$v_{\rm X-ray} \sim 450^{+192}_{-143}$\,km\,s$^{-1}$. A putative
SMBH binary would share this velocity. Projection effects are not 
known but if this velocity is purely in the plane of the sky, it
would have a secular proper motion $\mu_{\rm X-ray} \sim
5.7^{+2.4}_{-1.8}\,\mu$as\,yr$^{-1}$. ngVLA monitoring may reveal 
such a secular motion, upon which any reflex motion of 
$M_{\rm pri}$ would be superposed. Such a complication can be 
handled as for Sgr\,A$^\star$, where proper motion studies
exclude any oscillatory reflex signals \citep{rei04}.

In contrast to numerous prior studies that aim to detect the cm
emission from both SMBHs in the remnant of a galaxy merger
\citep[][and references therein]{bur19,der19}, our strategy needs
detectable mm emission from only one SMBH. Also unlike those studies,
our strategy can leverage the frequency coverage, angular resolution,
and sensitivity of the ngVLA to conduct searches well into the regime
in which a SMBH binary emits GWs, a topic further developed in
Section~4.3.

\subsection{Tie-Ins to Galaxy Evolution}

Figure~2 shows the values of $a_{\rm bin}$ associated with fiducial
SMBH binary periods of $P = 4$\,yr and $P = 40$\,yr. Astrometric 
monitoring through a quarter of a period would be sufficient to 
constrain the range of mass ratios $q$ allowed for the period. As is
evident from Figure~2, orbits with these periods could be spatially 
resolved by the ngVLA.

If no reflex motion is detected for $M_{\rm pri}$ after 1\,yr of 
ngVLA monitoring of NGC\,4472, a SMBH binary with period $P = 4$\,yr
and mass ratio $q > 0.01$ could be excluded. The darker blue shading
in Figure~2 indicates where SMBH binaries with shorter periods and 
higher mass ratios could also be excluded. This exclusion is
based on ngVLA astrometry and applies even for orbits that are not
spatially resolved with the ngVLA.

If reflex motion remains undetected after a decade of ngVLA
monitoring of NGC\,4472, a SMBH binary with $P = 40$\,yr and 
$q > 0.003$ could be excluded. Shorter periods with higher mass 
ratios could also be excluded, as shown by the lighter blue shading
in Figure~2.

Below, we cast such constraints on $q$, the mass ratio of the SMBH
binary, in terms of traits traceable to galaxy merger events as 
NGC\,4472 is assembled. The close coupling between galaxy and SMBH 
masses makes it reasonable to assume that $q$ for a SMBH binary 
will be similar to the mass ratio of the two progenitor galaxies 
which, after merging, give rise to the SMBH binary. Major galaxy 
mergers involve progenitors of comparable mass, whereas minor 
galaxy mergers involve progenitors of dissimilar mass
\citep[e.g.,][]{vol03}. Following the common practice of separating 
major and minor mergers at a mass-ratio boundary of about 
$\frac{1}{4}$, Figure~2 implies access to both types of galaxy 
mergers.

\subsubsection{Large q Values}

We first examine whether large values for the mass ratio $q$ could be
relevant to NGC\,4472. \citet{ran19} use numerical simulations to show
that the dissipationless, major merger of two galaxies can yield a SMBH 
binary that builds a stellar core, with tangentially biased orbits,
that is counter rotating relative to the surrounding merger remnant,
a massive early-type galaxy. 

While not meant to model NGC\,4472 specifically, the simulations of
\citet{ran19} do appear to offer a potential explanation for the
kinematics of its stellar core (see Section~2). In this context, 
ngVLA astrometric monitoring of NGC\,4472 would be seeking 
present-day evidence for the SMBH binary at high $q$ and on 
sub-parsec scales (Figure~2). 

A dissipationless scouring process shrinks the binary orbit on a 
timescale $t_{\rm sc} \propto a_{\rm bin}^{-1}$, slow compared to 
the timescale $t_{\rm gw} \propto a_{\rm bin}^{+4}$ for orbit 
shrinkage driven by GW emission 
\citep[e.g., Supplementary Figure~5;][]{min17}. 
Stated differently, the residence time over which a SMBH binary 
might be detected via astrometry is longer than the residence time
over which it might be detected via GW emission.

\subsubsection{Small q Values}

Small $q$ values could also be relevant to NGC\,4472, given 
observational evidence suggesting it has built up some of its size
and mass by accreting satellite galaxies \citep[e.g.,][]{jan10,arr12}.
As noted by \citet{kor13}, typical merger trees involve galaxies that 
could transport their own SMBHs inward. For example, an in-coming SMBH
with a mass $M_{\rm sec} \sim 2 \times 10^7$\,M$_{\odot}$ could lead
to a SMBH binary with a mass ratio $q \sim 0.01$, accessible
astrometrically according to Figure~2.

\subsection{Tie-Ins to GW Findings}

NGC\,4472 is sufficiently close that, if it hosts a SMBH binary, any
GW emissions from that binary might be detectable by current or
near-future pulsar timing arrays. As such, there is the possibility to
obtain independent checks on certain parameters or more comprehensive
information about the system than would be available from only
electromagnetic (ngVLA) or gravitational wave (pulsar timing)
measurements.

ngVLA astrometric monitoring could independently check the GW-based 
upper limits on $q$ plotted in Figure~2 for NGC\,4472, provided the 
GW findings invoke the same distance and SMBH binary mass adopted 
from \citet{sch16} for the astrometry. For the PPTA, the $q$ values
in Figure~2 are as tabulated by, and thus consistent with, the
\citet{sch16} distance and SMBH binary mass.

For NANOGrav, we focus first on a GW frequency of~8\,nanoHz (nHz), at
which the most constraining limits on GW strain can be set
\citep{agg19,arz21}. From equation~(1) of \citet{sch16}, we set a 95\%
upper limit on the associated chirp mass of $M_{\rm chi} < 0.44 \times
10^9$\,M$_{\odot}$ for our adopted distance to NGC\,4472.  We then
insert the ratio of the chirp mass to our adopted binary mass into
equation (5) of \citet{sch16} to set the 95\% upper limit of $q <
0.064$ that we plot in Figure~2. These upper limits on $M_{\rm chi}$
and $q$ are higher than those shown near 8\,nHz in Figure~2 of
\citet{arz21}; such differences arise because that study adopts a
cosmic-flow distance which is larger than the \citet{sch16} distance.

The NANOGrav search spans a GW frequency range of 2.8\,nHz to
317.8\,nHz \citep{agg19}. This range corresponds to orbital periods of
a putative SMBH binary from $P = 22.6$\,yr to $P = 0.2$\,yr,
equivalent to $a_{\rm bin} = 10600$\,au to $a_{\rm bin} = 453$\,au for
$M_{\rm bin} = 2.4 \times 10^9$\,M$_{\odot}$ \citep{sch16}.
GW-derived separations are so small that they are often expressed not
in pc, but in au. (An au axis is provided in Figure~2.) The $PSF$
adopted for the ngVLA astrometry of NGC\,4472 corresponds to 1670\,au,
making it complementary to and midway within the range of separations
constrained by \citet{agg19}. Those GW constraints degrade
significantly below about 5\,nHz, due to the 11\,yr data span. Future
observations, potentially enhanced with ngVLA pulsar timing data
\citep{cha18}, will improve the NANOGrav constraints, and extend them
to lower GW frequencies or longer orbital periods.

Finally, most GW analyses have been conducted assuming circular
orbits. As might be expected, for a SMBH binary on an elliptical
orbit, GWs are emitted not only at a frequency determined by the
orbital period but also at higher harmonics. The consequence is that
there can be a penalty in signal-to-noise ratio if a circular orbit is
assumed, but the orbit is elliptical \citep{hue15,tay16}. This
penalty particularly applies if the orbital period is shorter than
$1/T$, where $T$ is the data span for the pulsar timing. Thus, the
limits quoted above for NGC\,4472 become less constraining if one
considers elliptical orbits, particularly those at the higher GW
frequencies probed by NANOGrav. If ngVLA astrometry of NGC\,4472 were
to suggest that it contained a SMBH binary on an elliptical orbit,
improved limits on the chirp mass~$M_{\rm chi}$ or mass ratio~$q$
could likely be obtained.

\section{Summary and Conclusions}

A direct EM signature of an SMBH binary would be to spatially 
resolve and monitor the orbit of one or both of its members. To 
that end, we adapted the strategy of \citet{saf19} for EHT 
targets at 230\,GHz to the more abundant targets observable with
the ngVLA at 80\,GHz. The strategy involves astrometric monitoring
to trace the reflex motion of the binary's primary SMBH as it is 
tugged on by the secondary SMBH. Only one member of the binary 
needs to have detectable mm emission.

Picking the early-type galaxy NGC\,4472 as an example and casting
the circular orbit of a putative SMBH binary in terms of its 
period $P$, semimajor axis $a_{\rm bin}$, and secondary-to-primary
mass ratio $q$, we found the following:

\begin{enumerate}
\item Orbits with fiducial periods of $P = 4$\,yr and 40\,yr
could be spatially resolved and monitored with the ngVLA.
\item For a 95\% accuracy of $2\,\mu$as per ngVLA monitoring epoch, 
sub-parsec values of $a_{\rm bin}$ could be accessed over a range of 
mass ratios notionally encompassing major ($q > \frac{1}{4}$) and 
minor ($q < \frac{1}{4}$) galaxy mergers.
\item If no reflex motion was detected for the primary after 
1(10)\,yr of ngVLA monitoring, a SMBH binary with period 
$P = 4(40)$\,yr and mass ratio $q > 0.01(0.003)$ could be excluded.
This would suggest no present-day evidence for a past major merger 
like one recently simulated, wherein scouring by a $q \sim 1$ SMBH
binary formed a stellar core with the kinematic properties of 
NGC\,4472's stellar core.
\item Astrometric monitoring with the ngVLA could independently 
check the upper limits on $q$ for NGC\,4472 from searches for 
continuous gravitational waves with the PPTA and NANOGrav.
\end{enumerate}

\acknowledgments

We thank the reviewer for their helpful and timely report.

The NRAO is a facility of the National Science Foundation (NSF),
operated under cooperative agreement by Associated Universities,
Inc.\ (AUI). The ngVLA is a design and development project of the NSF
operated under cooperative agreement by AUI.

The NANOGrav project receives support from NSF Physics Frontiers
Center award number 1430284. Part of this research was carried out at
the Jet Propulsion Laboratory, California Institute of Technology,
under a contract with the National Aeronautics and Space
Administration.

This paper makes use of the following ALMA data: ADS/JAO.ALMA\#
2015.1.00926.S. ALMA is a partnership of ESO (representing its member
states), NSF (USA) and NINS (Japan), together with NRC (Canada), MOST
and ASIAA (Taiwan), and KASI (Republic of Korea), in cooperation with
the Republic of Chile. The Joint ALMA Observatory is operated by ESO,
AUI/NRAO and NAOJ.

\software{astropy \citep{ast18}, CARTA \citep{wan20}}

\end{document}